\documentstyle[preprint,prl,aps]{revtex}
\begin{document}
\draft
\flushbottom

\title{Quasiparticle spectrum of grain
boundaries in d-wave superconductors}
\author{M.J. Calder\'on$^{1}$ and E. Bascones$^{1,2}$}
\address{$^{1}$ Departamento de Teor\'{\i}a de la 
Materia Condensada, Instituto de Ciencia de
Materiales de Madrid (C.S.I.C.), \\ 28049 Cantoblanco, Madrid, Spain.\\
$^{2}$ Departamento de F\'isica de la Materia Condensada, Facultad de Ciencias, Universidad Aut\'onoma de Madrid, \\ 28049 Cantoblanco, Madrid, Spain. }
\date{\today}
\maketitle
\begin{abstract}
A grain boundary which separates domains with different orientation
along the c-axis is analyzed.     
The coupling of two parallel superconducting
planes whose order parameters are
rotated leads to interesting properties
which are expected to be common to other
kinds of grain boundary. The density of states
is enhanced at low energies. 
The number of localized states depends on the degree of misfit across the
boundary. A continuum of zero energy states appears at certain values of
the angle of rotation. 
\end{abstract}
\pacs{}

High temperature superconductors have been deeply studied 
since their discovery in 1986\cite{BM}. 
These new materials are based on planar 
copper oxide structures that are insulating and present a
phase transition to a superconducting state when they are doped.
The microscopic models proposed\cite{DA} to explain the appearance
of superconductivity in these materials point to a non conventional
symmetry of the order parameter. 
Now, it seems that finally the d-wave model is fully accepted 
\cite{exp1,exp2,exp3,exp4,exp5,exp6,tricrystal,tricrystal2,Tsuei2,Sun2,MAKI,VanH} as the 
correct one, at least for hole-doped
superconductors. 
The determination of the pairing symmetry has been possible thanks to the
experiments done using different techniques. 
Those experiments done in samples with grain boundaries have been
especially important. It is well known that lattice
distortions, such as grain boundaries, provide an intrinsic source of 
frustration in anisotropic superconductors because the order parameter follows the
orientation of the lattice axes. These systems have been widely studied in the
literature both experimentally\cite{tricrystal,tricrystal2,Tsuei2,Sun2,Sun1,Chaud93,Chaud94}
 and \cite{paco1,paco2,RICE,Kallin,Sigrist2,ZHITO,ScalPRB,paco3,Belzig}. 
 First experimental results were
controversial. Experiments reported by Sun and collaborators \cite{Sun1} showed
the observation of a small but finite Josephson coupling between a conventional 
s-wave superconductor and a heavily twinned film of YBCO. This was interpreted as
an evidence against d-wave symmetry. Josephson coupling across interfaces between
two YBCO regions with crystalline axis misoriented by $45^{o}$ led Chaudhari and
Lin\cite{Chaud94} to the same conclusion. On the other hand, the detection of a half quantum
of flux on tricrystal ring experiment\cite{tricrystal,tricrystal2} gave a clear 
evidence of the existence of nodes and lobes in the gap, supporting
the d-wave model. Recently it has been reported an experiment\cite{Tsuei2} which, based
on symmetry considerations, demostrate a pure d-wave order parameter in the 
tetragonal superconductor Tl2201.

However, the orthorhombic structure of YBCO allows for the existence of
$d + s$ pairing symmetry. Direct 
evidence of this behavior has been found on the magnetic interference 
pattern of a sample with a single twin boundary\cite{Sun2}, at which 
the direction of a and b axes reverse. As
the junction is formed in the superconducting plane                          
the whole system lacks periodicity. This fact introduces 
complexity in the study of the excitation spectrum. Latter
has been studied\cite{ZHITO} within the framework of the Bogoliubov-de Gennes (BdG)
equations and from ad hoc microscopic models.
The excitation spectrum presents a zero-energy peak in the local density of states near the 
twinning plane. Here we show that a finite density of states at
$E=0$ 
can also be obtained in another kind of grain boundaries.

In the present article, a new type of grain boundary is 
analyzed, see Fig. 1. We consider two parallel superconducting planes 
with d-wave pairing symmetry.
One of the planes has its order parameter rotated with respect to
the other by an angle $\phi_{0}$. 
Both planes are coupled  
by a hopping term t, which allows the
exchange of electrons between the two planes. Interplane tunneling
of Cooper pairs\cite{Anderson} could also be included in the model system but,   
for the sake of simplicity, we have not taken it into account. Notice
an important difference between our system, where the grain boundary is
formed between two planes, and grain boundaries usually considered in the
literature with only one plane involved.        
Due to the symmetry of our system, the spin
and momentum are conserved in the hopping process, contrary to what 
happens in  other kind of grain boundaries, where translational invariance
is violated.  
This fact opens us the possibility to obtain
the whole spectrum and to study the characteristic
features of grain boundaries avoiding quasiclassical approximations\cite{Maki2}. Despite of
its simplicity, we expect the spectrum of our system to reproduce the same 
behavior than in plane grain boundaries. Besides, this kind of structure could 
not only be found in YBCO but also in other compounds. In particular, the case 
$\phi_{o}=\frac{\pi}{2}$ is equivalent to $\pi$-junctions\cite{RICE,Sigrist2,ScalPRB}. 
It has also been suggested the
possibility that this type of structures is formed in granular materials as
well as during the growth process of the superconducting planes. A more realistic
model should consider the coupling of two semiinfinite media but we expect that the main
properties are captured by the present model.

We start from the mean field Hamiltonian            
\begin{eqnarray}
 H &  = &\sum_{k\sigma}\xi_{k}c^{\dagger}_{k\sigma\alpha}c_{k\sigma\alpha}
 + \sum_{k\sigma}\xi_{k}c^{\dagger}_{k\sigma\beta}c_{k\sigma\beta}\nonumber\\ 
&  + &\sum_{k}\Delta_{k\alpha}c^{\dagger}_{k\sigma\alpha}c^{\dagger}_{-k-\sigma\alpha}
+ \sum_{k}\Delta_{k\alpha}^{*}c_{-k-\sigma\alpha}c_{k\sigma\alpha}\nonumber\\
& + &\sum_{k}\Delta_{k\beta}c^{\dagger}_{k\sigma\beta}c^{\dagger}_{-k-\sigma\beta}
+ \sum_{k}\Delta_{k\beta}^{*}c_{-k-\sigma\beta}c_{k\sigma\beta}\nonumber\\
& + & \sum_{k\sigma}tc^{\dagger}_{k\sigma\alpha}c_{k\sigma\beta}
+ \sum_{k\sigma}tc^{\dagger}_{k\sigma\beta}c_{k\sigma\alpha}
\end{eqnarray}
where greek indices label the planes,
$c^{\dagger}_{k\sigma\alpha}$ creates an electron
with momentum $k$ and spin $\sigma$ in the plane $\alpha$, and
$\xi_{k}=\frac{\hbar^{2} k^{2}}{2m} - \mu$ with $\mu$ being 
the chemical potential. This dispersion relation has been chosen
because of its simplicity but the generalization to one with
a different dependence on $k$ is straightforward.  
The angle of rotation $\phi_{o}$ comes into the problem through
the relative values of the order parameters:

\begin{eqnarray}
 \Delta_{\vec{k}\alpha} & = & \Delta \cos 2\phi \nonumber\\
 \Delta_{\vec{k}\beta} & = & \Delta \cos 2(\phi+\phi_{o})
\end{eqnarray}
To calculate the eigenvalues of this Hamiltonian
the BdG equations are solved. After a straightforward
calculation we obtain:
\begin{eqnarray}
 E_{k}^{2} & = & \xi_{k}^{2}+t^{2}+
 \frac{\Delta^{2}}{2}[\cos^{2}2\phi+\cos^{2}2(\phi+\phi_{0})]\nonumber\\
 & \pm &
 \{\frac{\Delta^{4}}{4}[\cos^{2}2\phi-\cos^{2}2(\phi+\phi_{0})]^{2}
 +4\xi_{k}^{2}t^{2}\nonumber\\& + & t^{2}\Delta^{2}[\cos2\phi-\cos2(\phi+\phi_{0})]^{2}\}^{\frac{1}{2}} 
\label{espectro}
\end{eqnarray}
This spectrum presents states at the Fermi level. These zero energy states,
or nodes, are given by the equation:

\begin{eqnarray}
 \xi_{k}^{2} & = & t^{2}-\frac{\Delta^{2}}{2}
 [\cos^{2}2\phi+\cos^{2}2(\phi+\phi_{0})]\nonumber\\
 & \pm & \Delta\mid\cos2\phi+\cos2(\phi+\phi_{0})\mid\nonumber\\\
 & \times & 
 \sqrt{\frac{\Delta^{2}}{4}[\cos2\phi-\cos2(\phi+\phi_{0})]^{2}-t^{2}}
\label{nodes}
\end{eqnarray}
From this equation it is clear that nodes can only
exist if one of the following conditions is satisfied:
\begin{equation}
 \mid \cos2\phi-\cos2(\phi+\phi_{0}) \mid \geq \mid \frac{2t}{\Delta}\mid 
\label{condition1}
\end{equation}
or
\begin{equation}
 \cos2\phi+\cos2(\phi+\phi_{0})=0
\label{condition2}
\end{equation}
Two parameters are relevant: the hopping $t$ between the planes and the
angle $\phi_{o}$. For small values of $t$ the most relevant condition is
(\ref{condition1}) except for $\phi_{0}=\pi/2$. 
On the other hand, for sufficiently large values of $t$, 
(\ref{condition1})
is never fulfilled and the nodes are given by
\begin{equation}
 \xi_{k}^{2}=t^{2}-\Delta^{2}
 \cos^{2}2\phi
\label{exacta}
\end{equation}
with $\phi$ given by (\ref{condition2}).
The solution of eq.(\ref{nodes}) in the first Brillouin zone
is plotted in Figs. $2a$ and $2b$ for $\phi_{0}=\pi/6$ where
the influence of the hopping can be seen.
The nodes appear in pairs. As the hopping
increases, the distance between the nodes in a 
pair decreases. 
When (\ref{condition1}) cannot be satisfied
the nodes are aligned.
The situation is different 
for $\phi_{0}=\pi/2$ where a continuous line
of nodes is obtained. 
It can be seen that as the hopping increases more
values of $\phi$ satisfy (\ref{exacta})
and when $t\geq\Delta$ there are nodes for any $\phi$.
This behavior is shown in  Figs. $2c$ and $2d$.
The continuous line of nodes implies the existence 
of a finite density of states at $E=0$. This is
only observed for values of $\phi_{0}$ very close to
$\pi/2$.
In a superconductor with tetragonal symmetry
values of $\phi_{o}>\pi/4$ correspond to non-equilibrium situations.
However,
they can appear if a current is applied, as well as in orthorhombic 
superconductors with non-equivalent axes, where similar features are expected.

The density of states shows a characteristic multi-peak
structure due to van Hove singularities, which are associated to 
saddle points of $E_{k}$ given by eq (\ref{espectro}). It is worthy
to note that we have started from a $\xi_{k}$ which presents no
saddle points. As can be 
seen in Fig. $3$, the width of the peaks increases
with the hopping reflecting the splitting of the Fermi surface.
For high energies the density of states is a constant
as in normal metals in 2D.

In general there are localized states at the grain boundary. 
Notice that we use
localized states in the sense that for a given momentum there exist
excitations with an energy $E_{k}$ lower than the corresponding
excitation energy 
in absence of grain boundary ($\phi_{o}=0$).
\begin{equation}
 E_{k\gamma} = \sqrt{(\xi_{k}\pm t)^{2} + \mid \Delta_{\vec{k}\gamma}\mid^{2}}
\end{equation}
where $\gamma$ labels the index of the plane. 
The region of the Brillouin zone which presents
localized states is displayed in Fig. $4$. 
These states are mainly found inside the splitted 
Fermi surface ($\xi_{k}=\pm t$) for the unfrustrated case and
close to the values of $\phi$ which correspond
to the nodes. The area of this region increases
with $t$ and $\phi_{o}$.
This kind of grain boundary in a 3D sample could be described
as two coupled semiinfinite media. The localized states 
could not propagate into the bulk, but they would only exist at the
boundary. The effect of the semiinfinite media can be introduced in terms 
of a self-energy. As a consequence, the extended states will have a
finite mean lifetime. Note that, in a real sample, the planes are coupled
in groups. In some cases it is only necessary to consider the coupling inside the groups.
Thus, our results could be directly applied to double-layered
systems.

In conclusion, we have studied a new type of grain boundary, which
can appear in granular superconductors and, in general, during the growth
process. The excitation spectrum displays new interesting features
which seem to be common to other grain boundaries. Our model system
can be found in any superconductor with d-wave symmetry and not only
in the ones with orthorhombic structures as the widely studied 
twin boundaries\cite{Sun2,Sun1,Kallin,Sigrist2,ZHITO,ScalPRB}  
in YBCO. The simplicity of our model allows us to explore the whole
spectrum. In  more complex systems where
not only two planes but two semiinfinite media are coupled, the spectrum
is expected to display similar properties at the boundary. The quasiparticle spectrum
has nodes such that the positions of the latter 
depend on the coupling constant between the planes
and on the relative angle of the order parameters.    
The multi-peak structure of the density of states reflects the saddle points of 
the quasiparticle spectrum, while the enhancement at low energies is due to the 
existence of localized states.

We gratefully acknowledge the participants of the 'II Training Course in
the Physics of Correlated Systems and High Tc Superconductors' where this
work was initiated, in particular, to C. Batista, R. Killian, S. Di Matteos and
K. Maki. We also thank  G. G\'omez-Santos and T. Strohm for useful discussions.
We are particularly indebted to F. Guinea for his continuous encouragement,
helpful discussions and for reading the paper.
One of us (MJC) thanks support from Fundaci\'on Ram\'on Areces.

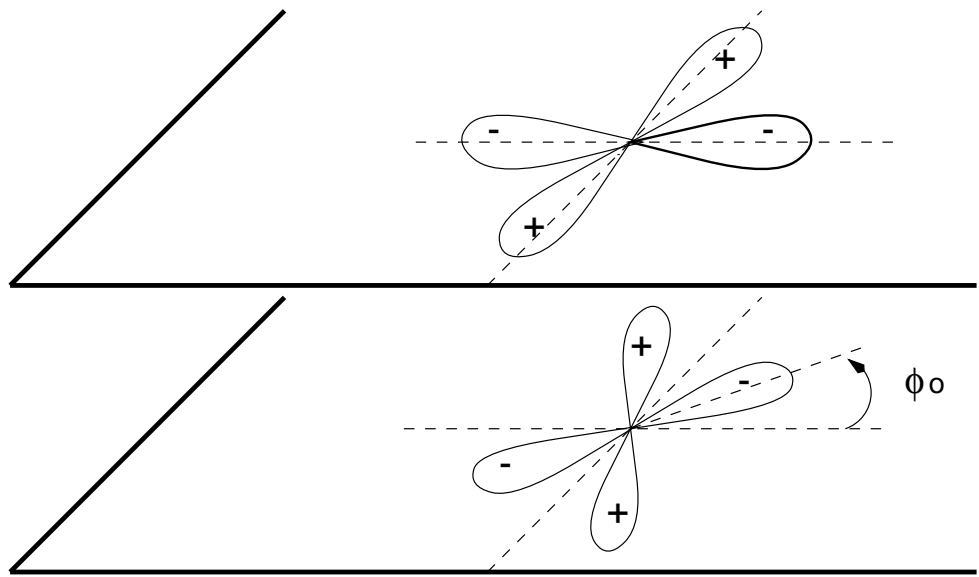
\begin{figure}
\caption
{
Sketch of our model system
}
\label{fig1}
\end{figure}

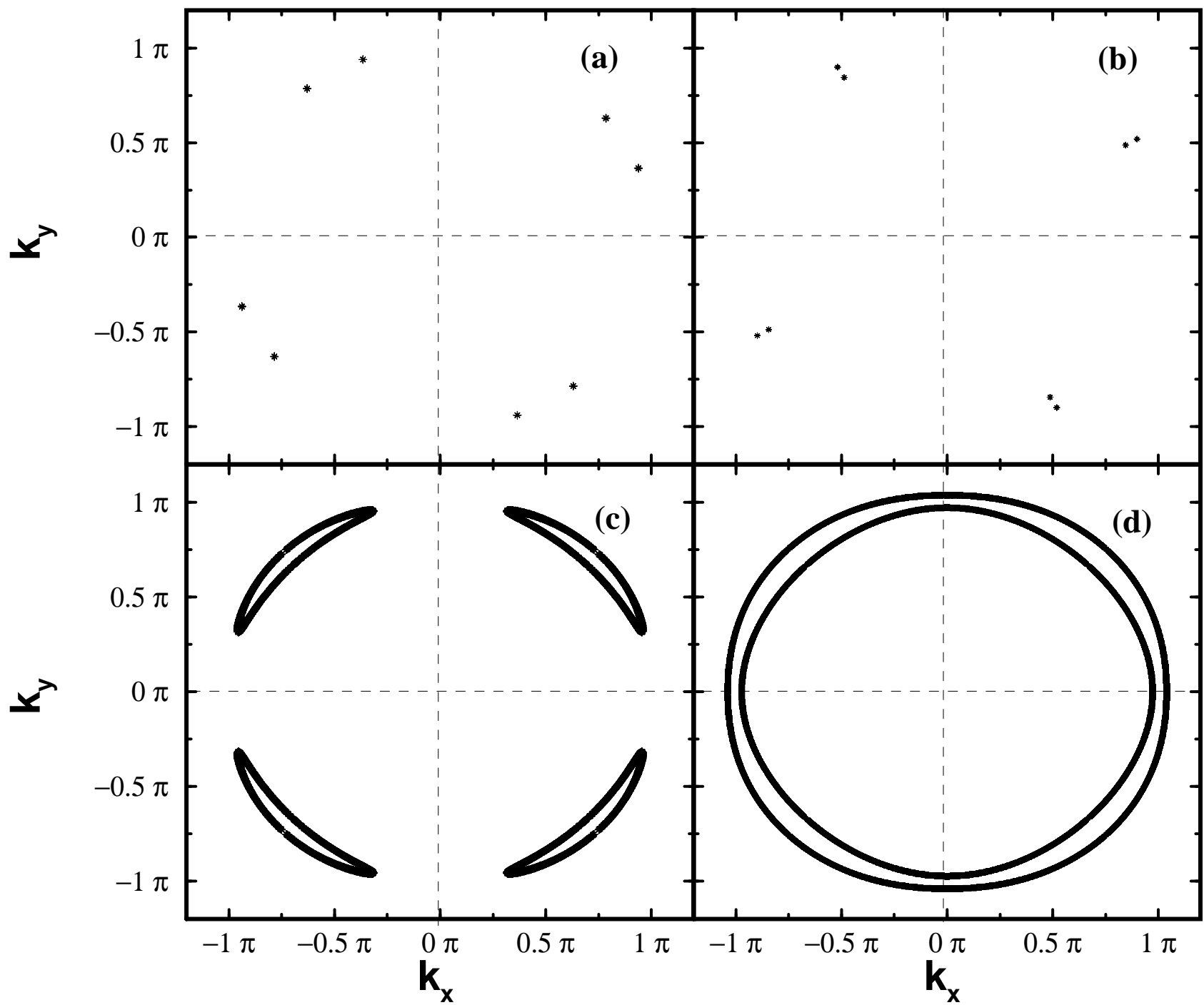
\begin{figure}
\caption
{
The nodes of the spectrum for $\phi_{0}=\frac{\pi}{6}$, $t=0.4\Delta$ in Fig. $2a$ and
$t=0.8 \Delta$ in Fig. $2b$,
and for $\phi_{0}=\frac{\pi}{2}$, $t=0.8 \Delta$ in Fig. $2c$ and $t=1.2 \Delta$ 
in Fig. $2d$, are plotted. 
$\phi_{0}=\frac{\pi}{2}$ displays an unusual behavior with a
continuum line of nodes in contrast to eight nodes found in almost all
the other cases. The influence of the hopping can also be seen. For
$\phi_{0}=\frac{\pi}{6}$ the nodes in a pair tend to be closer as the
hopping increases (see text). For $\phi_{0}=\frac{\pi}{2}$ as the hopping
grows an increasing number of nodes is obtained. Here, $k$ is given in 
units of $a^{-1}$, where $a$ is the lattice constant. Finally, 
$\frac{\hbar^{2}\pi^{2}}{2ma^{2}}= 40 \Delta$ and $\mu = 10 \Delta$.
}
\label{fig2}
\end{figure}

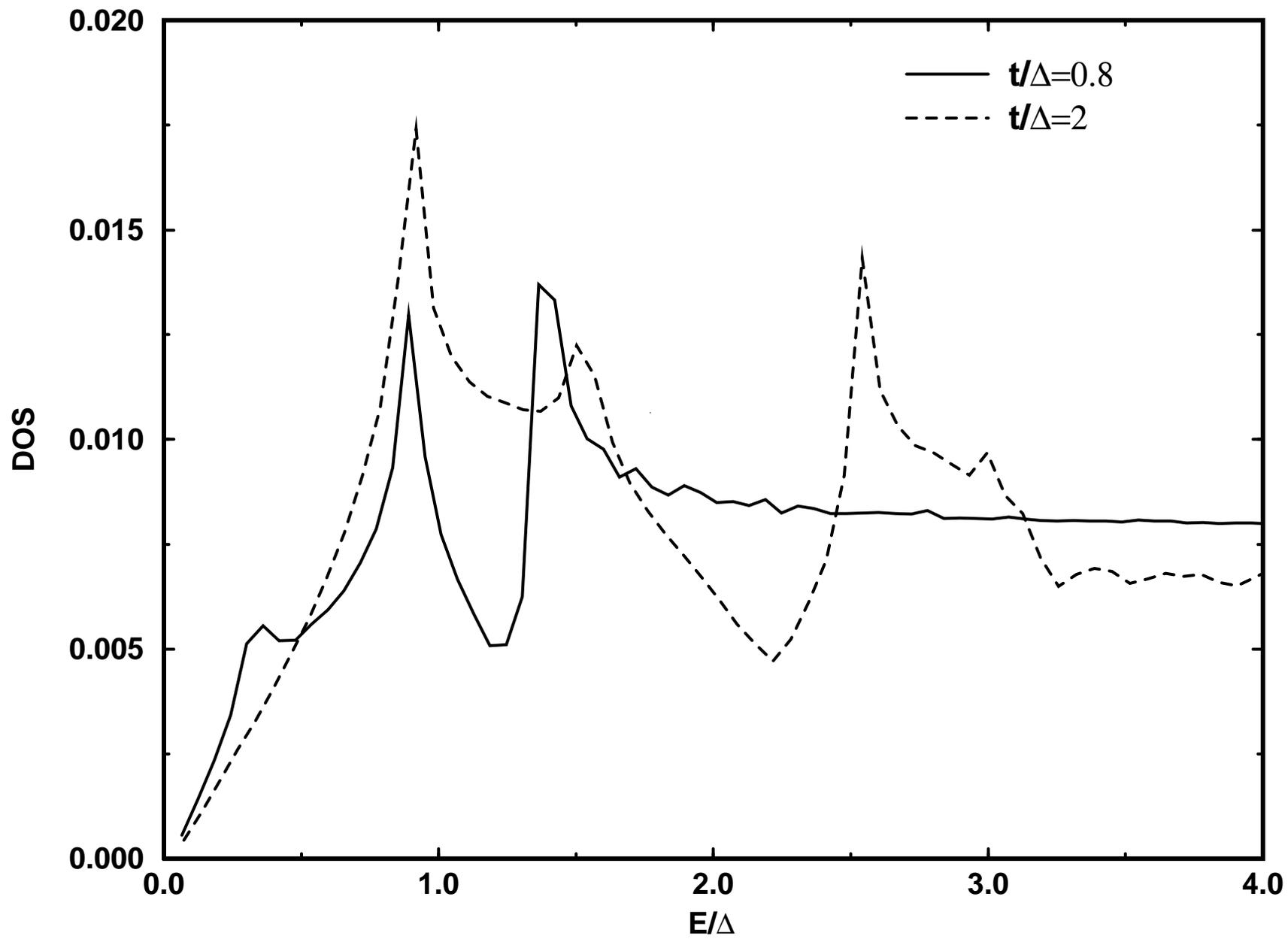
\begin{figure}
\caption
{
The density of states for $\phi_{0}=\frac{\pi}{6}$, $t=0.8\Delta$ and $t=2\Delta$
is plotted. The enhancement of the density of states at low energies 
increases with the hopping. The hopping also influences the width and number
of peaks which appear.
}
\label{fig3}
\end{figure}

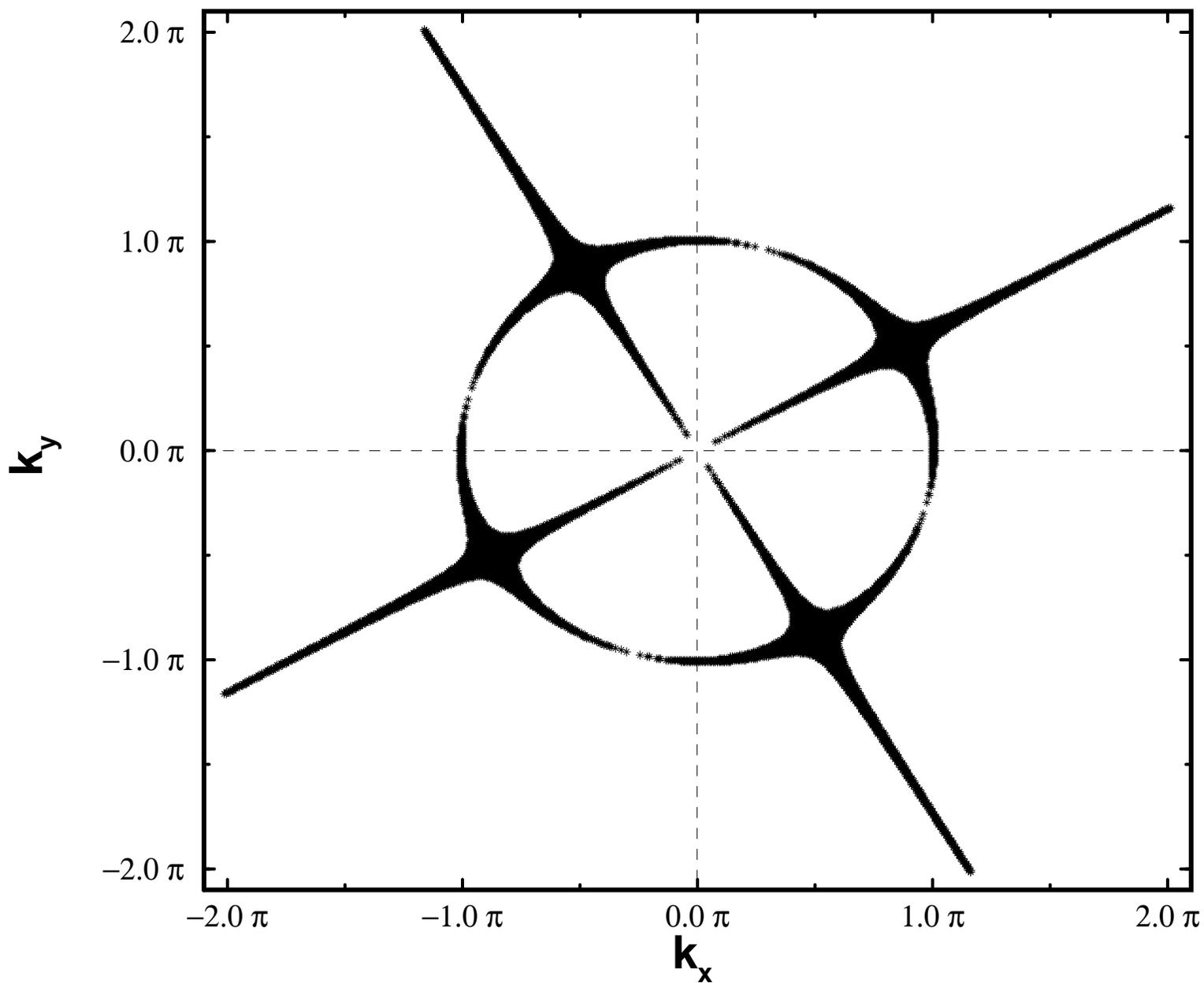
\begin{figure}
\caption
{
The figure shows the region of the Brillouin zone (BZ) which 
displays localized states for the case $\phi_{0}=\frac{\pi}{6}$ and $t=0.8\Delta$.
Units of $k$ and values at the boundaries of BZ are the same as those 
given in the caption of Fig. 2  
}
\label{fig4}
\end{figure}                     
\end{document}